\documentstyle[epsfig]{mn}

\newif\ifAMStwofonts

\def\apm{APM~0827+5255}
\def\b{$b$}
\def\cfour{C{\sc \,iv}}	
\def\cs{$\chi^2$}
\def\er{Equation~\ref}
\def\fr{Figure~\ref}
\def\kmps{km s$^{-1}$}
\def\logn{$\log N$}

\def\lya{Lyman $\alpha$}
\def\mgtwo{Mg{\sc \,ii}}
\def\pcmsq{cm$^{-2}$}
\def\vp{{\sc vpfit}}

\def\scr{Section~\ref}
\def\spose#1{\hbox to 0pt{#1\hss}} 
\def\tr{Table~\ref}
\def\approxlt{\mathrel{\spose{\lower 3pt\hbox{$\sim$}}
        \raise 2.0pt\hbox{$<$}}}
\def\z{$z$}
\def\zap{$z\approx$~}
\def\approxgt{\mathrel{\spose{\lower 3pt\hbox{$\sim$}}
        \raise 2.0pt\hbox{$>$}}}

\ifoldfss
  \ifCUPmtlplainloaded \else
    \NewTextAlphabet{textbfit} {cmbxti10} {}
    \NewTextAlphabet{textbfss} {cmssbx10} {}
    \NewMathAlphabet{mathbfit} {cmbxti10} {} 
    \NewMathAlphabet{mathbfss} {cmssbx10} {} 
  \fi
  \ifAMStwofonts
    \ifCUPmtlplainloaded \else
      \NewSymbolFont{upmath} {eurm10}
      \NewSymbolFont{AMSa} {msam10}
      \NewMathSymbol{\upi}     {0}{upmath}{19}
      \NewMathSymbol{\umu}     {0}{upmath}{16}
      \NewMathSymbol{\upartial}{0}{upmath}{40}
      \NewMathSymbol{\leqslant}{3}{AMSa}{36}
      \NewMathSymbol{\geqslant}{3}{AMSa}{3E}

       \let\ge=\geqslant
    \fi
  \fi
\fi 

\ifnfssone
  \newmathalphabet{\mathit}
  \addtoversion{normal}{\mathit}{cmr}{m}{it}
  \addtoversion{bold}{\mathit}{cmr}{bx}{it}
  \newmathalphabet{\mathbfit} 
  \addtoversion{normal}{\mathbfit}{cmr}{bx}{it}
  \addtoversion{bold}{\mathbfit}{cmr}{bx}{it}
  \newmathalphabet{\mathbfss} 
  \addtoversion{normal}{\mathbfss}{cmss}{bx}{n}
  \addtoversion{bold}{\mathbfss}{cmss}{bx}{n}
  \ifAMStwofonts
    \ifCUPmtlplainloaded \else
      \UseAMStwoboldmath
      \makeatletter
      \new@mathgroup\upmath@group
      \define@mathgroup\mv@normal\upmath@group{eur}{m}{n}
      \define@mathgroup\mv@bold\upmath@group{eur}{b}{n}
      \edef\UPM{\hexnumber\upmath@group}
      \new@mathgroup\amsa@group
      \define@mathgroup\mv@normal\amsa@group{msa}{m}{n}
      \define@mathgroup\mv@bold\amsa@group{msa}{m}{n}
      \edef\AMSa{\hexnumber\amsa@group}
      \makeatother
      \mathchardef\upi="0\UPM19
      \mathchardef\umu="0\UPM16
      \mathchardef\upartial="0\UPM40
      \mathchardef\leqslant="3\AMSa36
      \mathchardef\geqslant="3\AMSa3E

       \let\ge=\geqslant
    \fi
  \fi
\fi 

\ifnfsstwo
  \DeclareMathAlphabet{\mathbfit}{OT1}{cmr}{bx}{it}
  \SetMathAlphabet\mathbfit{bold}{OT1}{cmr}{bx}{it}
  \DeclareMathAlphabet{\mathbfss}{OT1}{cmss}{bx}{n}
  \SetMathAlphabet\mathbfss{bold}{OT1}{cmss}{bx}{n}
  \ifAMStwofonts
    \ifCUPmtlplainloaded \else
      \DeclareSymbolFont{UPM}{U}{eur}{m}{n}
      \SetSymbolFont{UPM}{bold}{U}{eur}{b}{n}
      \DeclareSymbolFont{AMSa}{U}{msa}{m}{n}
      \DeclareMathSymbol{\upi}{0}{UPM}{"19}
      \DeclareMathSymbol{\umu}{0}{UPM}{"16}
      \DeclareMathSymbol{\upartial}{0}{UPM}{"40}
      \DeclareMathSymbol{\leqslant}{3}{AMSa}{"36}
      \DeclareMathSymbol{\geqslant}{3}{AMSa}{"3E}

       \let\ge=\geqslant
    \fi
  \fi
\fi 

\ifCUPmtlplainloaded \else
  \ifAMStwofonts \else 
    \def\upi{\pi}
    \def\umu{\mu}
    \def\upartial{\partial}
  \fi
\fi

\title[Size estimates for \cfour\ absorbers]
{Size estimates for intervening \cfour\ absorbers from high resolution
spectroscopy of \apm.}

\author[Tzanavaris \& Carswell]
       {Panayiotis Tzanavaris\thanks{e-mail: pt2@ast.cam.ac.uk}, Robert F. Carswell\thanks{e-mail: rfc@ast.cam.ac.uk} \\
        Institute of Astronomy, Cambridge CB3 0HA}
\date{Accepted 
      Received;
      }

\pagerange{\pageref{firstpage}--\pageref{lastpage}}
\pubyear{2002}

\begin{document}

\maketitle

\label{firstpage}

\begin{abstract}
A new analysis of Keck/HIRES observations of the broad absorption line
QSO \apm\ indicates that a number of intervening \cfour\ absorbers
give rise to absorption lines for which the observed optical depths
for 1548, 1550~\AA\ doublet components are not in the expected $2:1$
ratio. To compensate for the effect, a local adjustment of the
zero-level is required. We model this effect as coverage of one line
of sight to this gravitationally lensed QSO and perform a set of
simulations to select a sample of lines for which our model provides
an explanation for the effect. We use lines in this sample to obtain
estimates for minimum \cfour\ absorber sizes from total coverage and
the separations of the lines of sight for a range of lens redshifts,
$z_{\rm l}$, and two cosmologies. We also obtain best estimates for
overall sizes from a statistical \lq hit and miss\rq\ approach. For
$z_{\rm l}=0.7$ our results set a lower limit to sizes of \cfour\
absorbers of $\sim 0.3 \, h^{-1}_{72}$ kpc ($\sim 0.5 \, h^{-1}_{72}$
kpc) for $\Omega_M=1, \Omega_{\Lambda}=0$ ($\Omega_M=0.3$,
$\Omega_{\Lambda}=0.7$), in agreement with other results from similar
work but are limited by sample size and the uncertainty in $z_{\rm
l}$. Our method can be used to detect lensed QSOs and to probe
absorber sizes when separate spectra cannot be obtained for each line
of sight.
\end{abstract}

\begin{keywords}
cosmology: observations -- gravitational lensing -- intergalactic medium --
methods: miscellaneous -- quasars: absorption lines --
quasars: individual (\apm)  
\end{keywords}

\section{Introduction}

The advent of 10-m class telescopes in the last decade has provided us
with an unprecedented view of the high-redshift universe. In
particular, HIRES, the high-resolution echelle spectrograph on the
Keck I telescope in Hawaii, has made it possible to obtain
observations with resolution as high as $\sim 6$~\kmps. As a result, a
number of detailed studies of absorption lines in the spectra of
background QSOs have been carried out (for reviews see Rauch 1998;
Hamann \& Ferland 1999).

However, this improvement in {\em spectral} resolution cannot be
easily accompanied by an improvement in {\em spatial} resolution:
2-dimensional information about the absorbers can normally only be
obtained by comparing separate spectra of multiple quasar images,
which are due to either gravitationally lensed single QSOs or actual
QSO pairs which have a small angular separation on the plane of the
sky. Because lensed QSOs have image separations up to a few
arcseconds, they are useful for probing scales $\approxlt 100$~kpc. QSO
pairs have image separations of at least a few arcminutes, thus
allowing one to probe scales of a few hundred kpc. From coincident
absorption in more than one line of sight (LOS), minimum size
(coherence length) estimates can be obtained immediately. Estimates of
most probable sizes require a statistical \lq hit and miss\rq\
approach (e.g. McGill 1990).

Such studies have led to size estimates for \lya\ absorbers ranging
from a few kpc (Foltz et al. 1984; McGuill 1990; Bechtold \& Yee 1995)
to a few hundred kpc (Dinshaw et al. 1995; Smette et al. 1995; Dinshaw
et al. 1998) to $\sim 1$~Mpc (Fang et al. 1996; Dinshaw et al. 1997;
Young et al. 2001). Such large \lq sizes\rq\ suggest that the typical, low
column density ($\approxlt 10^{14.5}$\pcmsq) \lya\ forest line does not
originate in discrete galactic or protogalactic clouds. The emerging
picture is consistent with results from hydrodynamical numerical
simulations in a Cold Dark Matter (CDM) structure formation scenario
according to which the \lya\ forest is a manifestation of a space
filling photoionized intergalactic medium (IGM). This is the most
important baryon reservoir in the Universe at high redshift (see
e.g. Efstathiou, Schaye \& Theuns 2000 for a review).

The \cfour\ doublet ($\lambda \lambda 1548.20, 1550.78$~\AA) 
is the most commonly observed heavy-element
absorption signature redward of the \lya\ emission line in spectra of
high-redshift QSOs.  The oscillator strengths for the 1548.20 and
1550.78~\AA\ components are in a $2:1$ ratio, which translates
to the same optical depth ratio for optically thin ($\tau \ll 1$)
lines.

Line parameters are consistent with a photoionized gas (Rauch et
al. 1996).  The \cfour\ absorbers appear to be highly clustered
(Sargent et al. 1988; Petitjean \& Bergeron 1994; Loh, Quashnock \&
Stein 2001) and have been thought to originate in gas clouds orbiting
in galactic haloes (Sargent et al. 1979; Steidel 1993; Petitjean \&
Bergeron 1994; Mo \& Miralda-Escud{\'e} 1996). Hydrodynamical
simulations also support a picture in which high-redshift \cfour\
arises in protogalactic clumps (PGCs), i.e., small, future
constituents of large galaxies, forming through gravitational
instability in a hierarchical cosmogony (Haehnelt, Steinmetz \& Rauch 1996;
Rauch, Haehnelt \& Steinmetz 1997).

Size estimates for heavy element absorbers range from a few kpc (Young
et al. 1981; Smette et al. 1992; Monier, Turnshek \& Lupie 1998) to a
few tens of kpc (Crotts et al. 1994; Smette et al. 1995; Lopez, Hagen
\& Reimers 2000) or even hundreds of kpc (Shaver \& Robertson 1983;
Petitjean et al. 1998). A recent study using three gravitationally
lensed QSOs (Rauch, Sargent \& Barlow 2001) obtains size estimates on
the order of kiloparsecs. It is also apparent that sizes are
absorber-type dependent, with individual clouds for low-ionization
absorbers having dimensions on the order of a few $\times 10$~pc as
opposed to at least a few $\times 100$~pc for high-ionization
absorbers (Rauch, Sargent \& Barlow 1999; Rauch et al. 2001; Rigby,
Charlton \& Churchill 2002).

With an $R$-band magnitude of 15.2 (Irwin et al. 1998) \apm\ is one of
the intrinsically most luminous QSOs known (bolometric luminosity
$\approxlt 10^{15} L_{\odot}$, Egami et al. 2000; Ibata et
al. 1999). From molecular CO emission lines the estimated quasar
redshift is $z=3.911$ (Downes et al. 1999). It is thus well suited for
QSO absorption line studies. Additionally, the object is
gravitationally lensed (Ledoux et al. 1998; Ibata et al. 1999; Egami
et al. 2000) but the redshift of the lens is not known. As the
object is lensed into three images, it represents the first confirmed
odd-image lens system (Lewis et al. 2002). 

For this object it is not possible to obtain separate spectra for
different LOS with ground-based observations (image separation $\sim
0.38$~arcsec). However, if only one LOS is covered, there is an excess
in the continuum intensity that gets through to the observer.  If the
absorbing column is high enough, the observed optical depths of the
components of a heavy element doublet are not in the expected ratio,
providing a signature for this effect. We use the term Anomalous
Doublet Ratio (ADR) for such doublets. In such a case, in order to
obtain a satisfactory fit, one has to adjust the spectral zero level
locally.  A number of \mgtwo\ systems show this effect in \apm. By
means of equivalent width modelling Petitjean et al. (2000) have obtained
size estimates of $\sim 1$~kpc for \mgtwo\ cloudlets. From separate
near-infrared spectra for each LOS Kobayashi et al. (2002) obtain
upper limit estimates of $\sim 200$~pc for \mgtwo\ clouds in a damped
\lya\ system.

We have used the ADR to select lines from intervening \cfour\ systems
which may be due to coverage of one LOS towards \apm.  The
observations and data reduction are described in Section
\ref{sec:obs}, the fitting procedure in \scr{sec:fit} and our
partial coverage model in \scr{sec:model}.  We have run a set of
simulations to establish whether the effect can be genuine for each
fitted line, and these are described in \scr{simulation}. We have
then used a maximum likelihood analysis to obtain estimates for the
most probable sizes of the \cfour\ absorbers (Section
\ref{sec:size}). We have used the currently favoured cosmological
model with matter density $\Omega_M=0.3$, and vacuum energy density
$\Omega_{\Lambda}=0.7$ (Perlmutter et al. 1999; Riess et al. 1998;
Riess et al. 2001). To facilitate comparisons with previous work,
where $q_0=0.5 \ (\Omega_M=1, \Omega_{\Lambda}=0)$ is used almost
exclusively, we have also carried out the calculations for such a
model. $H_0=72 \ h_{72}$~\kmps\ Mpc$^{-1}$ (Freedman et al. 2001)
throughout. We discuss our results in \scr{sec:disc}
and conclude in \scr{sec:conc}.

\section[]{Observations and data reduction}\label{sec:obs}

The data used in this analysis were obtained through a programme of
high resolution observations on the 10-m Keck I telescope in Hawaii in
April and May 1998 using HIRES with the TeK 2048$\times$2048 CCD at
the Nasmyth focus. Details of the observations can be found in Ellison
et al. (1999b). The data have been made publicly available (Ellison et
al. 1999a) but we did not use the version available on the internet
because there are a number of discontinuities due to inappropriate
merging of the orders (also noticed by Petitjean et al. 2000).
Instead, we used the original data files. However, in two of the
exposures there was a spurious systematic shift in wavelength of $\sim
0.2$~\AA, due to incorrect dating and, hence, incorrect heliocentric
velocity correction. After correcting for this problem, we used Tom
Barlow's {\bf MA}una {\bf K}ea {\bf E}chelle {\bf E}xtraction
programme (see
http://www2.keck.hawaii.edu:3636/realpublic/inst/hires/m\linebreak
akeewww/index.html) to obtain a set of disjoint echelle orders,
wavelength calibrated and rebinned onto a linear wavelength scale of
0.04~\AA\ per pixel.

We have flux calibrated each echelle order individually by means of
the available standard stars and standard {\small IRAF} routines to
ensure that there were no discontinuities where they were joined.  The
orders were joined by means of specially written interactive {\small
IRAF} scripts which ensured a smooth joining of adjacent and
overlapping parts with weights proportional to the local
signal-to-noise ratio (SNR).  A continuous spectrum has thus been formed for
which the continuum does not change abruptly.  This continuum was
fitted using cubic spline functions in spectral regions deemed free of
absorption.

\begin{figure}
\begin{center} 
{\vskip-4mm}
\psfig{file=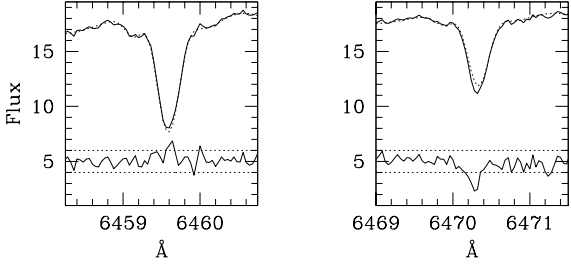, width=8cm}
\vspace{-4.0cm}
\end{center}
{\vskip-3mm}
\caption{Example of \cfour\ doublet showing an ADR.  The {\it left}
panel is for the 1548~\AA\ line and the {\it right} panel for the
1550~\AA\ line. In each panel from top to bottom shown are data ({\it
solid} line) and fit ({\it dotted} line), and residuals ({\it solid}
line) with $\pm 1\sigma$ level marked by two horizontal, {\it dotted}
lines. Flux units are arbitrary.}
\label{egzle}
\end{figure}
\section{Absorption Line Fitting}\label{sec:fit}

Apart from the 23 \cfour\ systems mentioned in Ellison et al. 1999b we
have fitted another 5 \cfour\ systems. We have detected one more
system which is rather uncertain and we have not used it in this
analysis.
Fitted redshift values range from 2.376647 to 3.671271. The latter
value corresponds to a radial velocity of $\sim -1.5\times 10^4$~\kmps\ 
with respect to the QSO. This reasonably ensures that the
corresponding absorbers are intervening.

The lines were fitted by means of the Voigt profile fitting
programme \vp\ (Webb 1987; Cooke 1994;
http://www.ast.cam.ac.uk/$\sim$rfc/vpfit.html) which uses $\chi^2$
minimization to produce lists of $b=\sqrt 2 \sigma$ (where $\sigma$ is
the standard deviation in a Gaussian distribution of velocities for
the absorbing atoms), column density, \logn, and redshift, $z$, for
all lines fitted.  Where the \cfour\ lines of interest were blended
with transitions at different redshifts all lines were fitted together
to produce a self-consistent fit. The wavelengths and oscillator
strengths used for the \cfour\ doublet are from Griesmann \& Kling
(2000). In some cases the lines of interest were blended with
atmospheric transitions which were identified by means of the standard
stars and an atmopheric line database. In order to obtain a
satisfactory fit in the region of interest, we fitted the atmospheric
transitions with Voigt profiles labelled \lq ??\rq, which were assigned the
\lya\ rest wavelength and oscillator strength.  This option was also
used in a few cases where it has not been possible to identify
absorption features with certainty. In all, we have fitted 100 \cfour\ 
doublets in the 28 systems which fall within the redshift interval mentioned above.

\begin{table*}
 \centering
 \begin{minipage}{120mm} 
  \caption{Lines for which there is need for a zero level
adjustment. Columns 7 and 8 indicate the adjustment, as a
fraction of the continuum, and the corresponding error,
respectively. Column 9 gives the covering factor calculated
from the continuum normalized residual intensities via
\er{equ:cfr}. The entries in Column 10 have been calculated directly from 
corresponding entries in Column 9. The index numbers refer to the full table of intervening
\cfour\ lines fitted as explained in the text. Lines whose index
numbers are prefixed by an x have not been used in this analysis
because the zero level adjustment does not correspond to the covering
factor of $\sim 0.4$ assumed in our model.}
  \begin{tabular}{@{}rcrrcccccc@{}}
\hline
   Index & $z$ & $b$ & $\sigma_b$ & $\log N$ & $\sigma_{\log N}$ & 
   Adjustment & Error & $f_R$ & $1-f_R$ \\[10pt]
\hline
1  &  3.108227  &  4.6  &  7.3  &  12.66  &  1.77  &  0.54  &  0.11  & 0.589547 & 0.410453\\
2  &  3.108299  &  12.8  &  18.0  &  13.07  &  1.31  &  0.54  &  0.11  & 0.652179 & 0.347821\\ 
3  &  3.108377  &  5.1  &  1.7  &  13.40  &  0.37  &  0.54  &  0.11  & 0.788434 & 0.211566\\ 
4  &  3.109511  &  10.8  &  1.4  &  13.45  &  0.06  &  0.54  &  0.11  & 0.580464 & 0.419536\\ 
13  &  3.134610  &  12.9  &  3.7  &  12.68  &  0.17  &  0.69  &  0.04  & 0.069223 & 0.930777\\
14  &  3.135135  &  16.1  &  2.0  &  12.96  &  0.07  &  0.69  &  0.04  & 0.082394 & 0.917606\\
15  &  3.133197  &  32.4  &  15.7  &  13.48  &  0.21  &  0.69  &  0.04 & 0.097442 & 0.902558\\ 
16  &  3.132676  &  13.3  &  4.4  &  12.94  &  0.31  &  0.69  &  0.04  &  0.097617 & 0.902383\\ 
17  &  3.133632  &  9.6  &  1.4  &  13.65  &  0.10  &  0.69  &  0.04  & 0.328090 & 0.671910\\ 
x15  &  3.172332  &  6.6  &  0.2  &  13.42  &  0.03  &  0.28  &  0.03  & 0.773285 & 0.226715\\ 
x28  &  3.202986  &  32.0  &  2.0  &  13.63  &  0.06  &  0.50  &  0.04 & 0.611514 & 0.388486\\ 
x29  &  3.202999  &  6.9  &  0.8  &  13.10  &  0.10  &  0.50  &  0.04 & 0.646416 & 0.353584\\ 

\hline
\label{tab:adj}
\end{tabular}
\end{minipage} 
\end{table*}
\section{Model}\label{sec:model}
\subsection{Motivation}
For a number of \cfour\ doublets the fit showed a tendency to be below
the data points for the 1548~\AA\ line and above the data points for
the 1550~\AA\ line. This was an indication that the optical depth 2:1
ratio of the model profile was not adequate for the particular
doublet, and it was a manifestation of an ADR, as mentioned above (see
\fr{egzle}).  The fitting parameters for such lines are shown in
\tr{tab:adj}. We stress that, short of including many extremely narrow
and unphysical lines ($b \ll$ instrumental resolution), the only way
to compensate for the effect is to adjust the zero level. In
particular, the effect cannot be attributed to continuum uncertainties
in the region of these lines. The zero level for a {\it given} line
which shows an ADR is effectively \lq stretched\rq\ by the addition of
a term, whose value is determined iteratively during the fitting
process. This does not affect the zero level of any other lines which
do not show an ADR, regardless of how close in velocity space these
may be to the adjusted line.  This adjustment can thus selectively
compensate for excess flux due to partial coverage from some, but not
all, absorbing clouds.

\subsection{Theoretical Considerations}
We now show how an ADR follows from relatively straightforward partial
coverage theory.

For a given true optical depth, $\tau_{\rm true} \propto \frac{N}{b}$,
when there is partial coverage, the observed intensity at a given
wavelength is given by
\begin{equation}\label{equ:i_cov}
I=(1-f)I_0+fI_0 e^{-\tau_{\rm true}} \equiv I_0 e^{-\tau_{\rm obs}}
\end{equation}
where $f\equiv\frac{{\rm Flux}_{\rm covered}}{{\rm Flux}_{\rm total}}$
is the covering factor, $I_0$ is the unabsorbed continuum intensity
and $\tau_{\rm obs}$ is the optical depth that corresponds to the
observed line profile. In what follows we shall use the label $F$ to
indicate a covering factor value obtained by means of {\it observed}
and {\it known} image {\it fluxes}.

It then follows that for a doublet 
\begin{equation}\label{equ:et1}
I_1=e^{-\tau_1}=1-f + f e^{-\tau_{\rm strong}}
\end{equation}
\begin{equation}\label{equ:et2}
I_2=e^{-\tau_2}=1-f + f e^{-\tau_{\rm weak}}
\end{equation}
where $\tau_{\rm strong}$ ($\tau_{\rm weak}$) is the true optical
depth of the strong (weak) component, $\tau_1$, $\tau_2$ are the
observed (apparent) optical depths and $I_1$, $I_2$ are the observed
residual intensities (normalized by the local continuum, $I_0$) at the
same velocity for the strong and weak component of the doublet,
respectively.

If the lines are optically thin, then from these exact expressions the
relation
\begin{equation}\label{equ:cfr}
f=\frac{1+I_2^2 -2I_2}{1+I_1 -2I_2}
\end{equation}
can be obtained which relates the covering factor to the observed
residual intensities (Hamann \& Ferland 1999). 
In what follows we shall use the label $R$ to indicate
a covering factor value obtained by means of {\it measured continuum
normalized residual intensities}.
All lines for which an
adjustment of the zero level has been performed with \vp\ are listed
in Table~\ref{tab:adj}.  The zero level adjustment, determined with
\vp, as a fraction of the continuum is given in column 7 and its
$1\sigma$ error in column 8. The covering factor determined from
\er{equ:cfr}, $f_R$, for these lines is given in column 9 of the Table. In
principle, since the zero level adjustment is introduced to compensate
for an excess continuum flux, the value for this adjustment should be
in agreement with $1-f_R$. The
last column in Table~\ref{tab:adj} shows the value of $1-f_R$ for all
entries, $f_R$, in column 9. In other words, the last column can be
thought of as giving an \lq expected\rq\ adjustment based on the
residual intensities.  If one then compares corresponding entries in
columns 7 (actual adjustment used) and 10 (expected adjustment from
residual intensities), one may notice that these agree to within
$1\sigma$ for line 17, to within $\approxgt 1\sigma$ for lines 1, 4
and 15, and to within $2\sigma$ or more for the rest of the lines. 
We call this an R-type comparison (see below) and we shall return to this
point later.

Further, from Equations~\ref{equ:et1} and~\ref{equ:et2} it is clear
that the ratio of the observed optical depths, $\tau_1/\tau_2$,
depends both on $f$ and on the true optical depth for each component,
and thus, for given $b$, on the column density. This {\it theoretical}
result is illustrated in \fr{fig:tauf} for a line with $b=10$~\kmps\
and different values of the column density and covering factor. It can
be seen that for low column densities and/or high covering factors,
this ratio is close to 2. This corresponds to the case where an ADR is
not observable.  However, for given $f$ and $b$ there is a threshold
in column density, above which the ratio will deviate significantly
from 2 and, depending on the SNR in the region of the line, it will
not be possible to fit the line with a model Voigt profile (which
assumes a ratio of 2) without obtaining large residuals. This effect
is seen in \fr{egzle}: relative to $\tau_2$, $\tau_1$ is less than
what \vp\ expects and, hence, the fit appears markedly below the
1548~\AA\ line and above the 1550~\AA\ line.
\begin{figure}
\begin{center}
\psfig{file=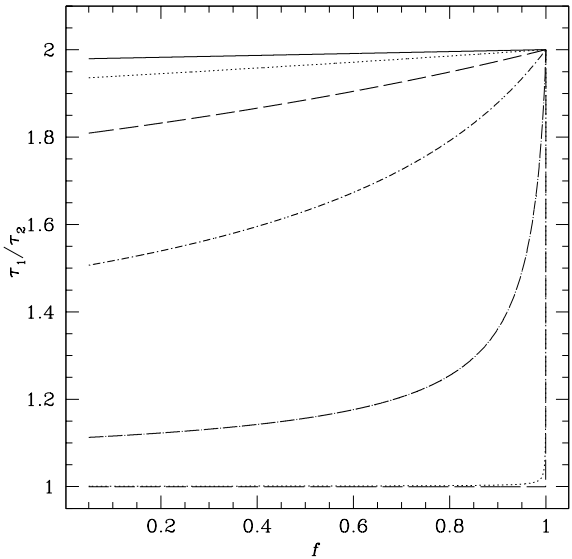, width=8cm}
\end{center}
\caption{Theoretical ADR. Shown here is the ratio of observed optical
depths, $\tau_1/\tau_2$ as a function of covering factor $f$ and
\logn(\cfour) for a \cfour\ line with $b=10$~\kmps. From top to
bottom, the curves are for \logn(\cfour) of 12.0, 12.5, 13.0, 13.5,
14.0, 14.5, 15.0.}
\label{fig:tauf}
\end{figure}

\subsection{Simulating an ADR}
This discrepancy is due to excess flux getting through at the velocity
where a particular doublet is observed. If the true optical depth and
the covering factor are known, the observed, continuum normalized
intensity is given by Equations~\ref{equ:et1} or~\ref{equ:et2}. Thus,
although the {\it true} continuum normalized profile is $e^{-\tau}$,
the profile actually {\it observed} will be
\begin{equation}\label{equ:add}
\frac{1-f}{f} + e^{-\tau} .
\end{equation}
This Equation is really the same as \er{equ:et1} (or \er{equ:et2}) but
with a different normalization.  In practice this means that, if one
wishes to simulate this effect for various values of the true optical
depth and covering factor, one needs to add a constant
$\alpha\equiv\frac{1-f}{f}$ to a continuum normalized spectrum. For
example, if there are two LOS of equal brightness and one is fully
covered while the other is uncovered, then $f=0.5$, $\alpha=1$ and if
one starts with a given, true absorption profile in a unit continuum,
the observable effects of partial coverage, if any, will be reproduced
if one adds a constant~$=1$ to all data points. The different
normalization of \er{equ:add} is convenient because excess flux can be
expressed in terms of a single additive constant, $\alpha$.

Thus a choice needs to be made on an appropriate value for $f$ (or,
equivalently, $\alpha$) since this is necessary for simulating an
ADR. In particular, in the case of \apm\ the most plausible lensing
model (Egami et al. 2000, Ibata et al. 1999), as well as recent
observations with the Hubble Space Telescope (Lewis et al. 2002),
indicate that the lensed system is made up of three images, A, B and
C, for which the flux ratio is A : B : C $=1 : 0.773 :
0.175$. However, image C lies closer to A, so A is less likely to be
the only one covered.  Therefore, it is plausible to assume that in
most cases there are, effectively, two images, A+C and B, either of
which, or both, may be covered by intervening absorbers.  Also, since
C contributes less than 10\% of the total flux, its presence (or
absence) will make little difference to the ADR.  There is also
further, more quantitative evidence that such a partial coverage
configuration is plausible. This was obtained as follows:

In \tr{tab:covs} we have tabulated all partial coverage combinations
which, from a geometric point of view, are reasonably probable. From
the known multiple image fluxes we calculated the covering factors,
$f_F$, and the corresponding \lq expected zero level adjustments\rq,
$1-f_F$, for each partial coverage configuration. We then compared the
latter with columns 7 and 8 ({\it actual} adjustment performed, $\pm$
error) of \tr{tab:adj} (we call this a comparison of \lq type F\rq).
This comparison suggests that, among the simple partial coverage
models in \tr{tab:covs}, the one with $\alpha=1.520$ ($f\sim 0.4$)
provides the best physical basis for most of the adjustments presented
in \tr{tab:adj}, except those whose index number is marked by an
x. Alternatively, it is possible to compare the last column of
\tr{tab:adj} (\lq expected zero level adjustment\rq, $1-f_R$, {\it
from the residual intensities}) to columns 7 and 8 of the same Table
(we call this a comparison of \lq type R\rq). However, for any given
line, $f_R$ is likely to be less reliable than $f_F$ as it depends on
the residual intensity measured, which is strongly affected by
blending due to other nearby lines. We have thus not based our choice
of model on an R-type comparison.  Nevertheless, it is interesting to
note that for line x15 which is only moderately blended, no simple
model among those shown in \tr{tab:covs} comes close to being
acceptable, on the basis of an F-comparison. However, this line is
only moderately blended, and an R-comparison shows that corresponding
values in columns 7 and 10 of \tr{tab:adj} {\it are} fairly close.
There is thus little doubt about the reality of the effect, but,
perhaps, there is a more complicated physical situation which requires
more sophisticated modelling.  On the other hand, model \lq A covered,
B+C uncovered\rq\ would be adequate for lines x28 and x29 (on the
basis of our chosen F-comparison).  In any case, we chose not to
discard lines for which there was poor agreement between columns 7 and
10 of \tr{tab:adj} (the R-comparison failed) as they might provide
useful insight after the simulations were complete.

To summarise, by performing Voigt profile fitting of intervening
\cfour\ lines, we have obtained a sample of 100 fitted \b--\logn\
pairs. For 12 of these lines an ADR was present and an adjustment of
the zero level has been applied in order that a good fit be
obtained. We have chosen a set of 4 possible partial coverage
configurations for this object and, using the known flux ratios of the
three images, have obtained 4 values for $1-f_F$, the \lq expected
zero level adjustment\rq. One of these values agrees best with the
greatest number of adjustments actually imposed on the 12 fitted
lines. We thus chose the corresponding model ($f\sim 0.4$) as the best
candidate for a physical explanation of the observed ADR.

\begin{table*}
 \centering \begin{minipage}{120mm} 
 \caption{Possible partial coverage
models for \apm.  From {\it left} to {\it right} columns give the
covering factor, $f_F$, calculated from the flux ratio A : B : C $=1
 : 0.773 : 0.175$, the value of $1-f_F$ which, ideally, should be in
good agreement with the adjustment of the zero level that needs to
be performed to obtain a good Voigt profile fit if an ADR is
observed, the image assumed covered, the image not covered and the
constant, $\alpha\equiv\frac{1-f}{f}$, that has to be added to a continuum 
normalized absorption spectrum to
simulate excess flux.}
  \begin{tabular}{@{}ccccc@{}}
\hline
   $f_F$ & $1-f_F$ & image covered & image not covered & $\alpha$ \\[10pt]
\hline
0.397 & 0.603 & B              & A+C &               1.520\\
0.487 & 0.513 & B+C            &  A  &               1.055\\
0.513 & 0.487 & A              & B+C &               0.948\\
0.603 & 0.397 & A+C            &  B  &               0.659\\
\hline
\label{tab:covs}
\end{tabular}
\end{minipage}
\end{table*}

\section{Simulations}
\label{simulation}

\subsection{Need for Simulations}
\label{sec:need}
A reliable way of establishing that the chosen partial coverage model
can (or cannot) give rise to a detectable ADR for a given \b--\logn--SNR
combination is to perform a set of simulations, as explained in the
next subsection.

Once the detectability, or not, of the ADR has been established, this
can be compared to the {\it actual} situation in the observed
spectrum.  For each \b--\logn--SNR combination there are four
possibilities:
\begin{enumerate}
\item ADR {\it not} detectable and {\it not} seen in observed spectrum,
\item ADR detectable {\it and} seen in observed spectrum,
\item ADR detectable but {\it not} seen in observed spectrum,
\item ADR {\it not} detectable {\it but} seen in observed spectrum.
\end{enumerate}
Cases (ii) and (iii) are useful for estimating absorber sizes: In case (ii)
{\it partial} coverage is established and, therefore, the transverse size of
the absorber must be {\it smaller} than the proper separation of the LOS at
the absorbing redshift. Conversely, in case (iii) {\it total} coverage is
established and the transverse size of the absorber must be {\it larger}
than the proper separation of the LOS.

Further, a maximum likelihood approach can be used to estimate most
probable absorber sizes from the number of established partial and
total coverage cases.

\subsection{Method}
Each of the 100 fitted \b--\logn\ pairs mentioned above
gives a single {\it true} $\tau$ value and can be used
to produce an artificial continuum normalized absorption
profile, $e^{-\tau}$. If one adds a term $\alpha=1.520$
to each data point in this profile, one effectively
introduces excess flux due to partial coverage from our
chosen physical model (\er{equ:add}).

Because the SNR varies considerably over the observed
spectrum from $\sim 25$ to $\sim 120$, for each region containing a
\cfour\ doublet component an average SNR value was determined by inspection
of the spectrum using the task {\it splot} in {\small IRAF}. 

For each \b--\logn(\cfour) pair two artificial spectra were produced.
The first spectrum contained 120 \cfour\ doublets with a profile
$e^{-\tau}$. The spacing of the lines was chosen empirically so that when subsequently
the lines would be fitted with \vp, blending complications would be absent, i.e. even
the broadest lines were well apart from each other.
The spectrum was resampled to the resolution of the observed spectrum
(0.04~\AA\ per pixel) and smoothed to the instrumental resolution of
6.6~\kmps. Gaussian noise was added with $\sigma=\frac{1}{\rm SNR}$ to
obtain the desirable final SNR, as measured beforehand for each region. 

The above process was repeated to produce another set of
120 simulations of the same line aiming to
reproduce the effects of excess flux.
This time gaussian noise with $\sigma=\frac{1+\alpha}{\rm SNR}$
was added to the unit continuum spectrum and both data and continuum were then
increased by $\alpha$. This ensured that the desired SNR was obtained
in this case as well.

Next, \vp\ was used to fit the artificial spectra. Each of the 120
artificial doublets in the spectrum was fitted independently, so that
120 separate \cs\ values were obtained. The starting \b--\logn\ values for
the \vp\ iterations were exactly the same as those used to make the
simulated lines, so that one would expect to obtain a good fit, unless the
effects of excess flux were significant.
\begin{table*}
 \centering
 \begin{minipage}{120mm} 
  \caption{Lines for which 
 simulations indicated that an ADR is detectable. 
 Columns 2 to 6 give the fitting parameters for the observed
 lines. Column 7 gives the significance for the KS statistic. The
 last column indicates whether the result means there is
 partial ({\it p}) or total ({\it t}) coverage.}
  \begin{tabular}{@{}rcrrcclc@{}}
\hline
   Index & $z$ & $b$ & $\sigma_b$ & $\log N$ & $\sigma_{\log N}$ & 
   Significance & Coverage \\[10pt]
\hline
  3 & 3.108377 &   5.1 &   1.7 & 13.40 &  0.37 & $< 0.001$ & {\it p}\\
  4 & 3.109511 &  10.8 &   1.4 & 13.45 &  0.06 & $< 0.001$ & {\it p}\\
  7 & 3.108564 &  18.6 &   1.6 & 13.53 &  0.08 & $< 0.001$ & {\it t}\\
 15 & 3.133197 &  32.4 &  15.7 & 13.48 &  0.21 & $< 0.04$ & {\it p}\\
 17 & 3.133632 &   9.6 &   1.4 & 13.65 &  0.10 & $< 0.001$ & {\it p}\\
 18 & 3.134133 &  12.6 &   0.8 & 13.14 &  0.05 & $< 0.05$ & {\it t}\\
 25 & 3.172054 &   6.3 &   1.1 & 12.25 &  0.06 & $< 0.05$ & {\it t}\\
 26 & 3.172580 &  17.5 &   3.2 & 12.57 &  0.08 & $< 0.04$ & {\it t}\\
 45 & 3.377585 &   9.8 &   0.3 & 13.20 &  0.02 & $< 0.001$ & {\it t}\\
 96 & 2.974079 &   9.1 &   0.4 & 13.19 &  0.02 & $< 0.03$ & {\it t}\\
\hline
\label{tab:sim0.4}
\end{tabular}
\end{minipage}
\end{table*}

\vp\ was allowed to fit only a single Voigt profile to each simulated line.
The option of automatically putting in additional narrow lines to
improve poor fits was not used since such lines were known to be
spurious by construction. The fitting region for each doublet component 
was centred on the component \lq redshift\rq\ and extended  
over
$3\times {\rm FWHM}$, where ${\rm FWHM}=2\sqrt{\ln 2} \ b$ is the
full width at half minimum. This was done to avoid including too much
continuum which can make a poor fit less pronounced.

If for a given \b--\logn\ pair the higher continuum introduced an
observable ADR, then, on average, the second set of 120 fits should be
markedly poorer than the first set. Otherwise, the two distributions
of \cs(reduced) obtained should not differ significantly.
\fr{fig:sim3} shows a single simulation of line 3 with the
Voigt profile fit superposed. An ADR was reproduced at $>2\sigma$ level
in the central region of the line. Note that a $2:1$ doublet ratio was
recovered for this simulated line when the zero level was adjusted by
$0.62\pm 0.02$, which agrees both with the original adjustment of
$0.54\pm 0.11$ (\tr{tab:adj}) and with $1-f_F=0.603$ for a simulated
line with $\alpha=1.520$ (\tr{tab:covs}).  Independently from this
result from \vp, the continuum normalized residual intensities gave
a covering factor $f_R=0.378$ ($1-f_R=0.622$), also in good agreement.
Additionally, the original Voigt parameters of the line were recovered
after the adjustment of the zero level ($b=5.0\pm 0.2$~\kmps, $\log
N=13.44\pm0.05$; compare with \tr{tab:adj}). This shows that this
type of adjustment successfully compensates for partial coverage
effects and that using fitted \b--\logn\ values to simulate lines with
an adjusted zero level is physically meaningful.

For each simulated \b--\logn\ pair we performed a two-sided Kolmogorov--Smirnov
(KS) test for the two cumulative distributions, $C_1$ and
$C_{1+\alpha}$ of \cs(reduced) obtained, in order to see if the
ADR had been reproduced, in a statistically significant sense, by the
simulations.  If the significance level for the KS statistic, defined
as $D\equiv \max \mid C_1(\chi^2)-C_{1+\alpha}(\chi^2) \mid$, was less
than 0.05, the null hypothesis that the two distributions were the same
was discarded.  This was interpreted as evidence that the second set of 120
fits was poorer and thus an ADR had been reproduced.  The significance
results for all such lines are given in Table~\ref{tab:sim0.4}. We
illustrate by plotting the cumulative distributions of the \cs\ values for fits to
simulations of line 3 in \fr{fig:ks3}.
\begin{figure}
\begin{center}
\psfig{file=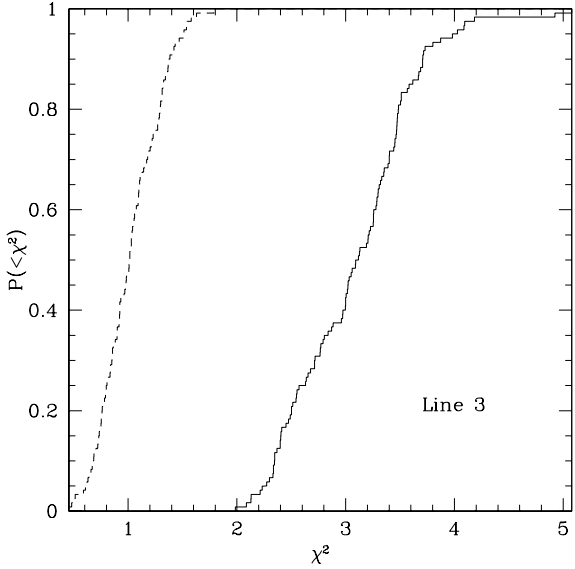, width=8cm}
\end{center}
\caption{\label{fig:ks3}Cumulative distribution of reduced \cs\ for
Voigt profile fits to simulated line 3. {\it Solid} line: distribution
of 120 \cs\ values from fits to simulated doublets with a continuum
$1+\alpha$. {\it Dashed} line: distribution of 120 \cs\ values from
fits to simulated doublets with a unit continuum.}
\end{figure}
\begin{figure}  
\begin{center}
\psfig{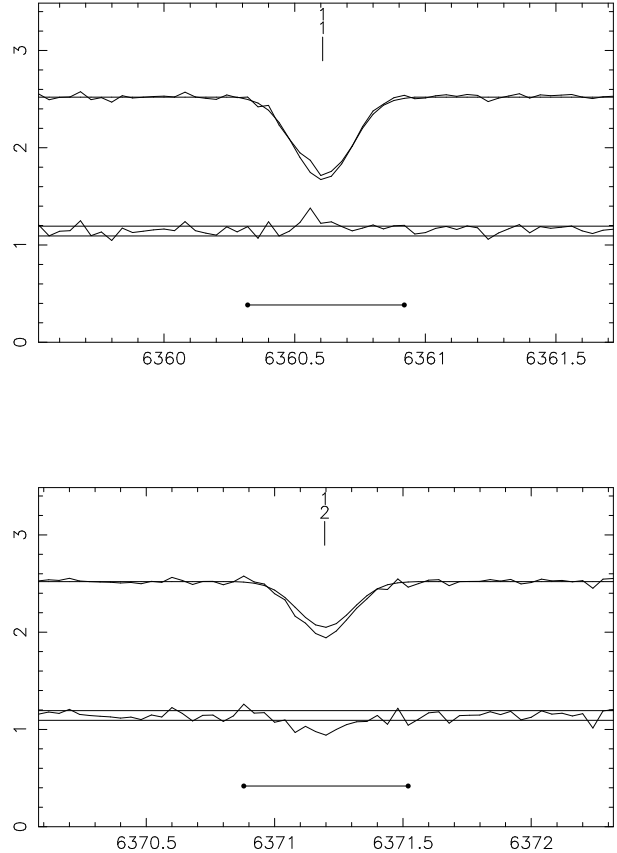}
\end{center}
\caption{\label{fig:sim3}Simulated \cfour\ doublet ({\it upper panel}:
1548~\AA\ region, {\it lower panel}: 1550~\AA\ region) showing an
anomalous doublet ratio. The Voigt parameters for the line are the
same as for line 3. Shown in each panel from top to bottom are data
and fit, residuals with $\pm 1\sigma$ level marked by the two
horizontal lines, and fitting region ($3\times$FWHM). Note how, as in
\fr{egzle}, near the line centre, the fit appears below (above) the
data in the 1548~\AA\ (1550~\AA) region.}
\end{figure}

If the corresponding originally fitted line also showed an ADR [case
(ii) in \scr{sec:need}], the assumed partial covering model was taken
to be the physical reason (a {\it p}-case. In work where separate
spectra are available for each LOS the term \lq anticoincidence\rq\ is
commonly used for such cases).  On the other hand, if the effect
appeared {\it only} in a simulation [case (iii) in \scr{sec:need}],
then, as far as this model was concerned, there was no partial but
total coverage (a {\it t}-case or \lq coincidence\rq\ in work with
separate spectra).

In \scr{sec:size} we explain how we used the number of {\it t} and
{\it p}-cases to obtain an estimate for the absorber sizes for two
distinct absorber geometries.

\subsection{Results}\label{sec:res}
According to the reasoning above, for $\alpha=1.520$ Table~\ref{tab:sim0.4} establishes 
the detectability of the ADR for the 10 simulated lines
listed there, of which 6 did not show the effect in the observed
spectrum (and were thus taken to be {\it t}-cases) and 4 also showed the
effect in the observed spectrum ({\it p}-cases).

\begin{figure}  
\begin{center}
\psfig{file=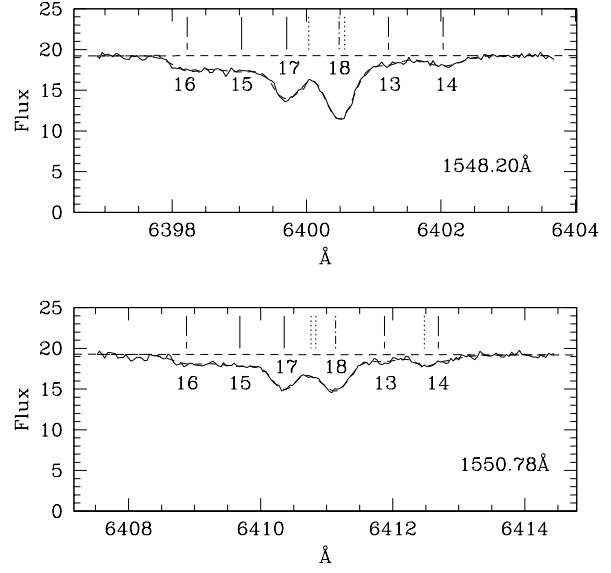, width=8cm}
\end{center}
\caption{As in \fr{fig:f3.10} but for the \zap3.13 system. Only in this Figure 
{\it dotted} ticks indicate unidentified/atmospheric lines.}
\label{fig:f3.13}
\end{figure}

\begin{figure}
\begin{center}
\psfig{file=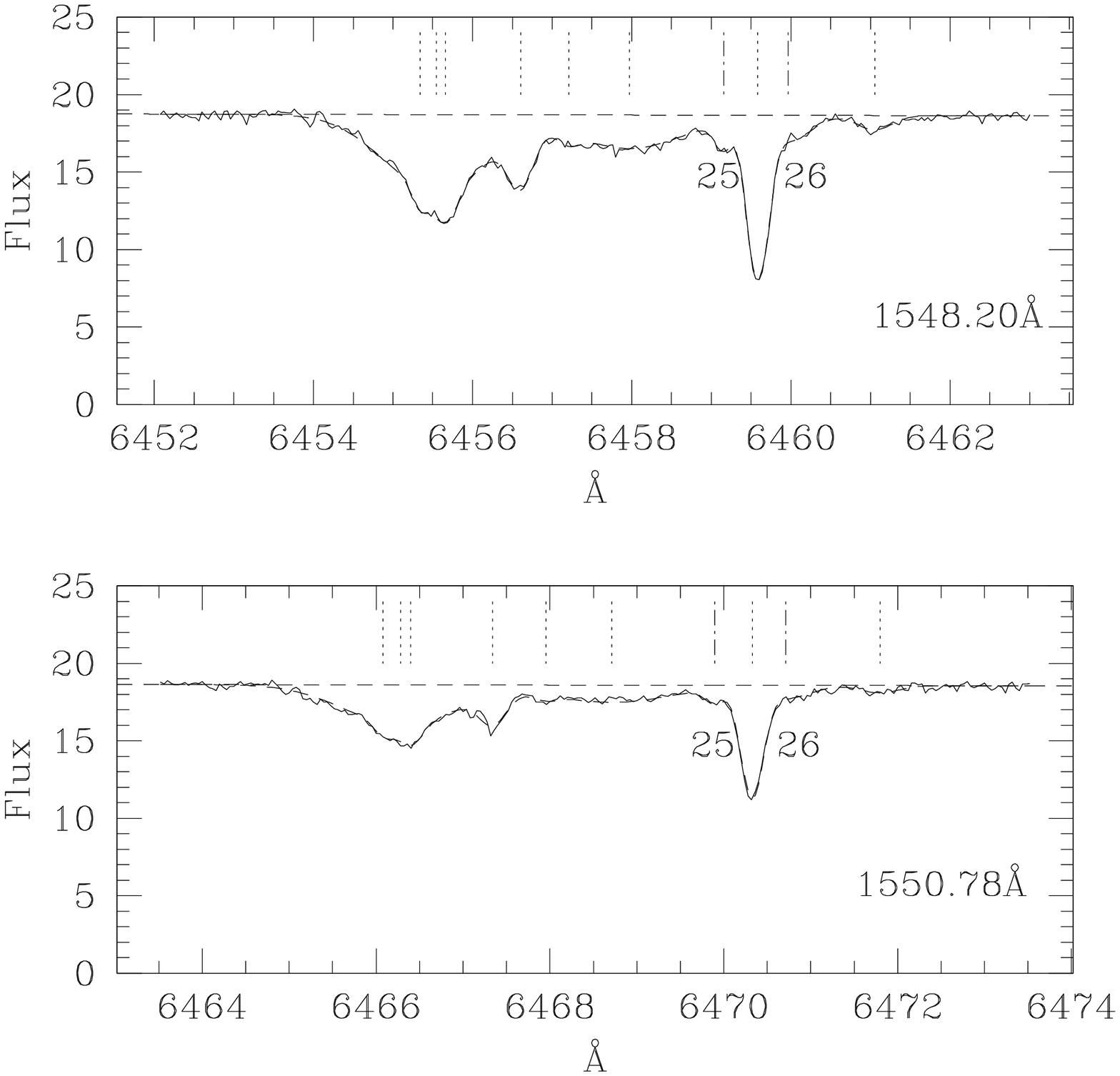, width=8cm}
\end{center}
\caption{As in \fr{fig:f3.10} but for the \zap3.17 system.}
\label{fig:f3.17}
\end{figure}

\begin{figure}
\begin{center}
\psfig{file=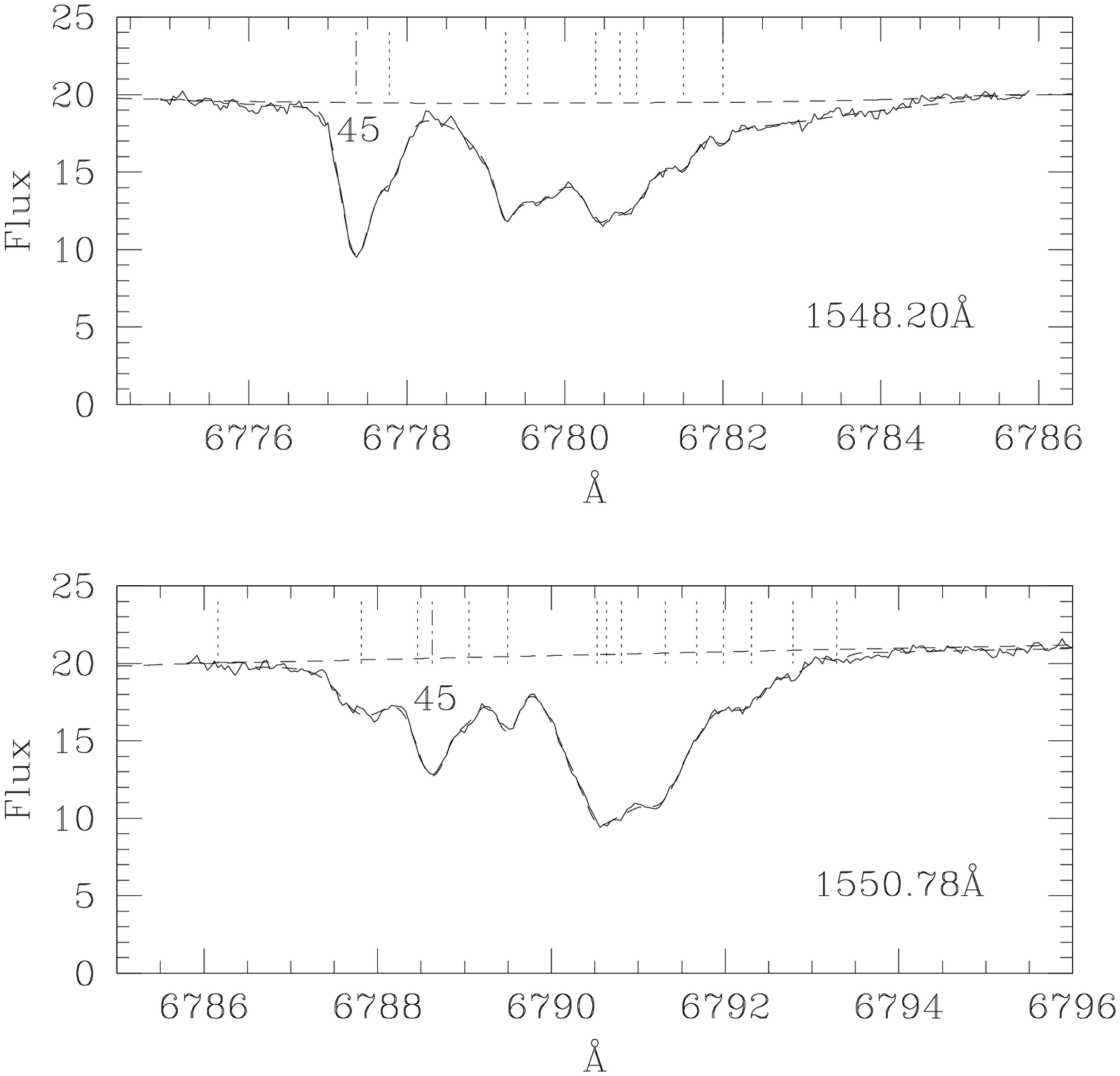, width=8cm}
\end{center}
\caption{As in \fr{fig:f3.10} but for the \zap3.37 system.}
\label{fig:f3.37}
\end{figure}

\begin{figure} 
\begin{center}
{\vskip-2.7mm}
\psfig{file=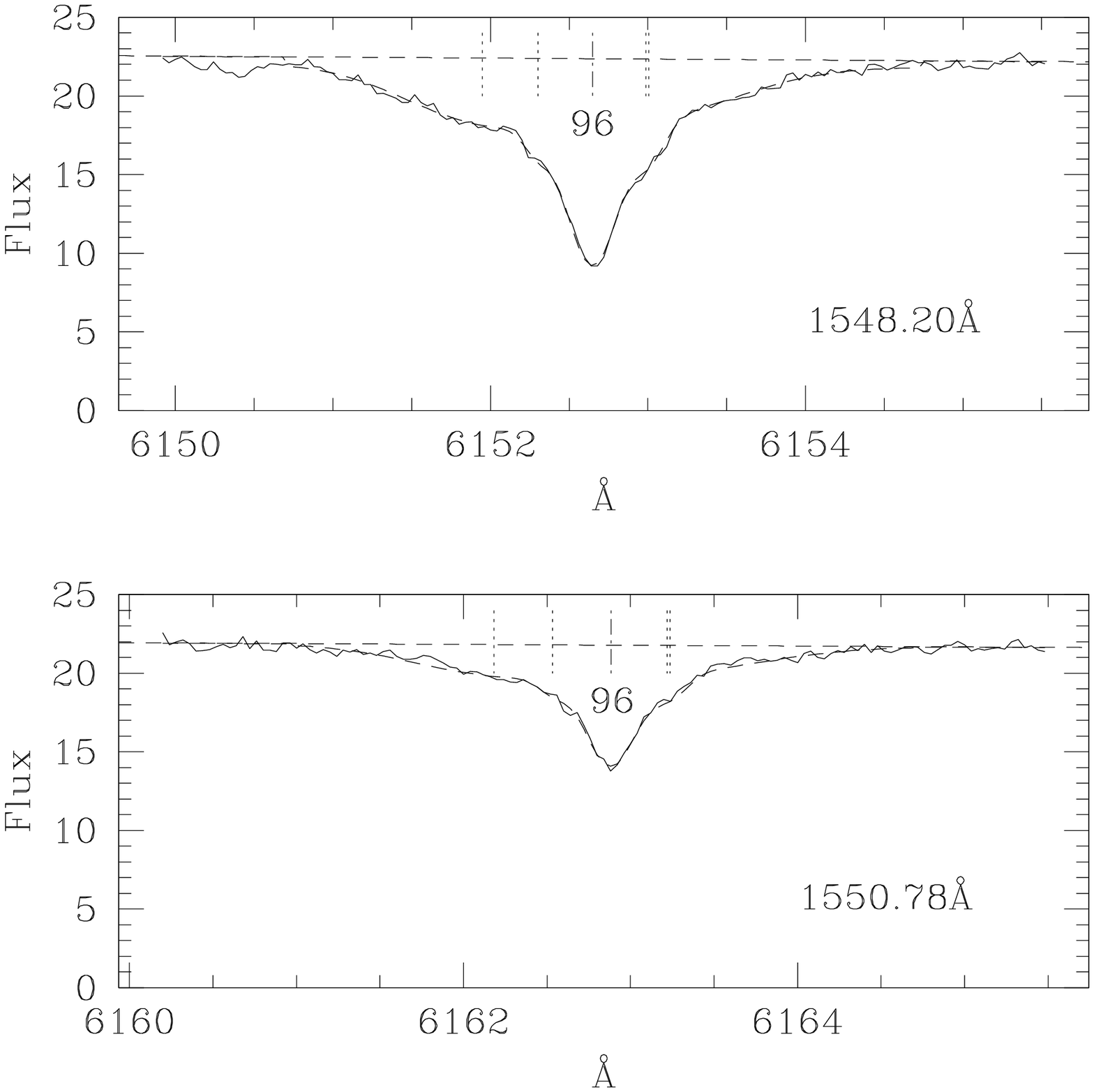, width=8cm}
\end{center}
\caption{As in \fr{fig:f3.10} but for the \zap2.97 system.}
\label{fig:f2.97}
{\vskip-2mm}
\end{figure}

More generally, all fitted and simulated lines can be divided into the four
groups introduced in \scr{sec:need}:

\begin{enumerate}
\item For the majority of lines no ADR was detected either during Voigt
profile fitting or in simulations. Such lines were not used further (but
see discussion).

\item For lines 3, 4, 15 and 17 the zero level has been adjusted during
fitting and the ADR was reproduced in simulations. These {\it p} lines are
indicated by {\it solid} ticks in Figures~\ref{fig:f3.10}
and~\ref{fig:f3.13}. These lines belong to each of the two systems
(\zap3.10 and \zap3.13) that show the effect {\it and}
were used in this analysis (see Table~\ref{tab:adj}).

\item For lines 7, 18, 25, 26, 45 and 96 an ADR was only produced
in simulations with varying degrees of significance.  Lines 7 and 18
are from the two systems at \zap3.10 and \zap3.13 for other lines of
which partial coverage was suggested by the simulations. Lines 25 and
26 belong to the \zap3.17 system in which line x15 shows an ADR which
cannot be modelled with $\alpha=1.520$. Lines 45 and 96 are from lower
redshift systems.  These {\it t} lines are indicated by {\it dot-short-dashed}
ticks in Figures~\ref{fig:f3.10} to~\ref{fig:f2.97}.

\item Finally, for lines 1, 2 (\zap3.10 system) and 13, 14, 16 (\zap3.13 system)
an ADR was not reproduced in simulations although they have been fitted
with a zero level adjustment.  These lines are indicated by {\it long-dashed}
ticks and numbers in Figures~\ref{fig:f3.10} and~\ref{fig:f3.13} and
were not used further.
\end{enumerate}

The existence of this last group of lines in a way provides an {\it a
posteriori} justification of simulations. To check the effect that
zero level adjustment of such lines had on the overall
goodness-of-fit, we modified the fit for the
\zap3.10 and 3.13 system as follows:

\begin{figure}  
\begin{center}
\psfig{file=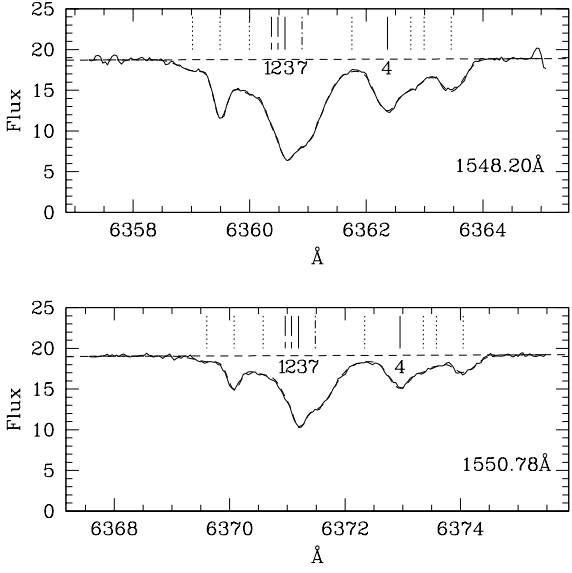, width=8cm}
\end{center}
\caption{Fitted absorption lines which may be due to partial or total
coverage of LOS to \apm\ as suggested by simulations for the \zap3.10
system. The fit is shown by a dashed line, superposed on the data.
The horizontal dashed line shows
the continuum. Numbers correspond to entries in Tables~\ref{tab:adj} 
and~\ref{tab:sim0.4}. 
{\it Solid} ticks and numbers indicate lines for which
the zero level has been adjusted at the Voigt profile fitting stage
(\tr{tab:adj}) and  for which an ADR was reproduced in the simulations for this model.
{\it Long-dashed} ticks and numbers indicate lines for which
the zero level has been adjusted at the Voigt profile fitting stage
but an ADR was not reproduced in the simulations.
{\it Dot-short-dashed} ticks and numbers indicate lines 
for which the zero level has not
been adjusted at the fitting stage but an ADR was reproduced in
simulations when excess flux was introduced (\tr{tab:sim0.4}). 
Other fitted \cfour\
lines are indicated by {\it dotted} ticks.}
\label{fig:f3.10}
\end{figure}

We used the last fitting results for each region as input to \vp\ but
excluded group (iv) lines from zero level adjustment. Otherwise, we did not
introduce any new lines. This procedure had a negligible effect on the
fit for the \zap3.10 system (reduced \cs\ changed from $\sim 1.09$ to $\sim
1.10$). The original adjustment of the zero level for lines 1 and 2 was
not necessary and was probably a result of the empirical nature of this
adjustment combined with the fact that these lines are close to the
stronger line 3. In the case of the \zap3.13 system and lines
13, 14 and 16 the situation is similar since the reduced \cs\ changed from
$\sim 1.16$ to $\sim 1.17$. 

The simulations thus allowed us to distinguish for which lines an ADR
may have a physical cause.

The choice of an upper limit for the significance of the KS statistic
was somewhat arbitrary. We chose 5\%\ in an effort not to discard too
many lines, since few lines were available anyway. However, the \cs\
cumulative distribution plots for lines 25 and 26
strongly indicate that a lower value might be
appropriate since
\cs\ values for fits of simulated lines with excess flux did not
show a clear trend to be higher. Since we were mainly interested in
illustrating a technique, we formally accepted the 5\%\ level and used
these two lines in further analysis. If this caveat is taken into
account, \tr{tab:sim0.4} suggests that a threshold column density for
an observable ADR in this model is given by
\logn(\cfour)~$\approxgt 13.1$ with $b\approx 10$~\kmps. This explains
the failure to reproduce lines in group (iv) above since all of these
lines (except, perhaps, for line 2, which, however, has large 
$\sigma_{{\rm log}N}$) are below this threshold.

\begin{table*}
 \centering
 \begin{minipage}{120mm} 
  \caption{Minimum sizes for \cfour\ absorbers. Columns 1 and 2 give
 the line index and fitted redshift for all coincidences established
 with simulations. Columns 3 to 7 give the separations of the two
 lines of sight for five different lensing redshifts. All separations
 are in $h_{72}^{-1}$ kpc. Here $q_0=0.5$ .}
  \begin{tabular}{@{}lcccccc@{}}
\hline
 Index & $z$ & $S(z_{\rm l}=0.7)$ & $S(z_{\rm l}=1.062)$ & $S(z_{\rm l}=1.1727)$
 & $S(z_{\rm l}=1.181)$ & $S(z_{\rm l}=2.974)$ \\[10pt]
\hline
7 & 3.108564 & 0.272 & 0.377 & 0.409 & 0.411 & 1.561\\
18 & 3.134133 & 0.262 & 0.363 & 0.394 & 0.396 & 1.504\\
25 & 3.172054 & 0.248 & 0.343 & 0.372 & 0.374 & 1.421\\
26 & 3.172580 & 0.248 & 0.343 & 0.372 & 0.374 & 1.420\\
45 & 3.377585 & 0.173 & 0.239 & 0.259 & 0.261 & 0.990\\
96 & 2.974079 & 0.326 & 0.451 & 0.489 & 0.492 & 1.868\\
\hline
\label{tab:sep}
\end{tabular}
\end{minipage}
\end{table*}
\begin{table*}
 \centering
 \begin{minipage}{120mm} 
  \caption{Same as Table~\ref{tab:sep} but for $\Omega_M=0.3$, $\Omega_{\Lambda}=0.7$.}
  \begin{tabular}{@{}lcccccc@{}}
\hline
 Index & $z$ & $S(z_{\rm l}=0.7)$ & $S(z_{\rm l}=1.062)$ & $S(z_{\rm l}=1.1727)$
 & $S(z_{\rm l}=1.181)$ & $S(z_{\rm l}=2.974)$ \\[10pt]
\hline
7 & 3.108564 & 0.362 & 0.518 & 0.566 & 0.570 & 2.317\\
18 & 3.134133 & 0.349 & 0.499 & 0.546 & 0.549 & 2.233\\
25 & 3.172054 & 0.330 & 0.471 & 0.516 & 0.519 & 2.110\\
26 & 3.172580 & 0.329 & 0.471 & 0.515 & 0.519 & 2.108\\
45 & 3.377585 & 0.230 & 0.329 & 0.359 & 0.362 & 1.471\\
96 & 2.974079 & 0.433 & 0.619 & 0.677 & 0.681 & 2.770\\
\hline
\label{tab:sep2}
\end{tabular}
\end{minipage}
\end{table*}

\section{Estimating Absorber Sizes}\label{sec:size}
\subsection{Direct Estimates}
Miminum size estimates can be obtained straightforwardly from lines
which correspond to total coverage.

The proper separation at reshift $z$ of the light paths from the two
images of a lensed QSO
with emission redshift $z_{\rm em}$ due to a lens at redshift $z_{\rm l}$ is given by
\begin{equation}
\label{sep}
S(z)=\frac {D(z_{\rm l},z_{\rm con}=0) D(z,z_{\rm em})} 
{D(z_{\rm l},z_{\rm em})} \delta\phi,
\end{equation}
(Young et al. 1981) where $\delta\phi$ is the observed angular separation of the
images. The light paths converge at $z_{\rm con}=0$ and the angular
diameter distances, $D$, are given by 
\begin{equation}
\label{angdiam}
D(z_1,z_2)=\frac{D_C(z_2)-D_C(z_1)}{1+z_2}
\end{equation}
where 
\begin{equation}
D_C(z)=\frac{c}{H_0}\int_0^z{\frac{dz'}{\sqrt{\Omega_M (1+z')^3 + \Omega_{\Lambda}}}}
\end{equation}
is the comoving distance for a flat universe ($\Omega_K=0$; Peebles
1993; Hogg 2000).  Equ.~\ref{sep} holds for $z>z_{\rm l}$ as assumed
here.  Egami et al. (2000) argue that the lensing galaxy
should be at $z_{\rm l} \sim 3$ based on the Einstein and core radii
of their model. The damped \lya\ system at $z\sim 2.974$ may be
compatible with this estimate. Additionally, Petitjean et
al. (2000) suggest that strong \mgtwo\ systems are prime lense
candidates because the roughly equal brightnesses of images A and B
suggest the LOS are traversing the central regions of the lensing
object.  We have thus calculated the separation for five values of
$z_{\rm l}$ as shown in Tables~\ref{tab:sep}
and~\ref{tab:sep2}.  These values include an arbitrary value of
0.7, not corresponding to any known object, three values corresponding
to observed \mgtwo\ systems and one value from the damped \lya\ system
mentioned.
The observed angular separation of image B from the combined A+C image
was taken to be $\delta\phi=0.369$~arcsec. There is some ambiguity as to what
one means by the location of the A+C image, and we have simply chosen
the location of the flux-weighted centroid of the A+C image. Size
estimates scale with $\delta\phi$ (this still holds, approximately, in
the statistical approach of the next section) and so the effect is of
the order of no more than a few percent. Thus the results remain
qualitatively unaffected.

\subsection{Maximum Likelihood Analysis}
We then applied a maximum likelihood method in order to estimate the most 
probable size of the absorbers, given the information that we have
obtained through this work. We assumed two simple geometries for
absorbers of uniform size. 

The probability that a {\it spherical} cloud is intersected by both
LOS, given that it is intersected by one is 
\begin{equation}
\label{phis}
\phi_s (X)=\frac{2}{\pi}\left\{ \arccos \left[ X(z)\right] - X(z)
\sqrt{1-X(z)^2} \right\}
\end{equation}
(McGill 1990), for $X \in [0,1]$, and zero otherwise. Here $X(z)\equiv 
S(z)/2R$, where $R$ is the absorber radius. For randomly inclined circular disks the
probability is
\begin{eqnarray}
\label{phid}
\lefteqn{ \phi_d (X)= \int_{-\pi/2}^{\pi/2} \frac{\cos\theta}{\pi}  }\nonumber \\
& & \times \left\{ \arccos \left[ \frac{X(z)}{\cos\theta} \right]
-\frac{X(z)}{\cos\theta} \sqrt{1-\frac{X(z)^2}{\cos^2\theta}} \right\} 
d\theta
\end{eqnarray}
for $X<\cos\theta$ and zero otherwise (McGill 1990). Here $\theta$ is
the angle of inclination of the disk relative to the lines of sight. 
We were actually interested in the case where there is absorption at
{\it either} LOS, so that the probability that
there is absorption at the other LOS is
\begin{equation}
\psi_{s,d}=\frac{\phi_{s,d}}{2-\phi_{s,d}}
\end{equation}
[by analogy to Dinshaw et al. (1997) who apply this to two separate
spectra]. 
The probability of getting the estimated number 
of cases where there is coverage of one or both lines of sight is then 
given by the likelihood function
\begin{equation}
\label{lik}
{\cal L}(R)=  \prod_i \psi_{s,d}\left[ X(z_i)\right] \prod_j \left\{1-\psi_{s,d}\left[
X(z_j)\right] \right\},
\end{equation}
where $i$ labels cases of total coverage ({\it t} lines) and $j$ cases of
single coverage ({\it p} lines). 

\begin{figure}
\begin{center}
{\vskip-4mm}
\psfig{file=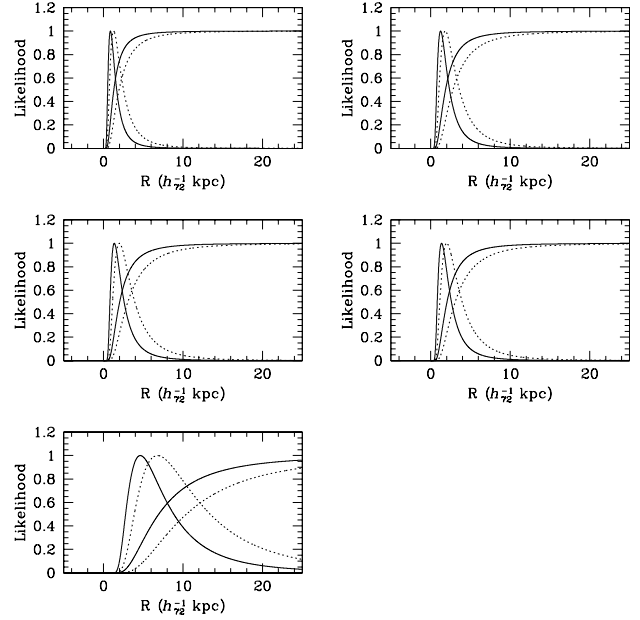, width=9cm}
\end{center}
{\vskip-3mm}
\caption{Differential and cumulative likelihood distributions for
spherical absorbers ({\em solid curves}) and randomly inclined disks
({\em dotted curves}) as functions of cloud radius in an
$\Omega_M=0.3$, $\Omega_{\Lambda}=0.7$ universe. The differential
distribution has been normalized to its peak value. From left to right
and top to bottom results are shown for $z_{\rm l}$ of 0.7, 1.062, 1.1727,
1.181, 2.974.}
\label{zlik}
\end{figure}

\begin{table*}
 \centering
 \begin{minipage}{120mm}
 \caption{Radius estimates of
 \cfour\ absorbers from maximum
 likelihood analysis in an $\Omega_M=1$, $\Omega_{\Lambda}=0$
 universe. 
 The most probable and median radii for spherical
 clouds and randomly inclined disks are shown for four different
 lensing redshifts. The 95\% confidence limits have been estimated from the
 cumulative distributions.}
 \begin{tabular}{@{}ccccccc@{}}
\hline
\hline
      & Spheres  &            &         & Disks   &            &\\
\hline
      &          &            & 95\% &            &            & 95\%\\
      &          &            &Confidence&        &            &Confidence\\
$z_{\rm l}$ & Mode $R$ & Median $R$ & Limits & Mode $R$ & Median $R$ & Limits\\
   &($h_{72}^{-1}$ kpc)&($h_{72}^{-1}$ kpc)&($h_{72}^{-1}$ kpc)
   &($h_{72}^{-1}$ kpc)&($h_{72}^{-1}$ kpc)&($h_{72}^{-1}$ kpc)\\
\hline
0.7000 &  0.650 &  0.980 &   0.380-4.330 &  0.960 &  1.480 &  0.540-6.900\\
1.0620 &  0.900 &   1.360 &    0.540-6.210 &  1.330 & 2.040 &  0.740-9.140\\
1.1727 &  0.980 &   1.470 &    0.570-6.470 &  1.440 & 2.210 & 0.790-9.850\\
1.1810 &  0.980 &   1.480 &    0.580-6.570 &  1.450 & 2.230 &  0.810-10.230\\
2.9740 &  3.120 &  4.680 &  1.810-20.150 &  4.570 & 7.030 &  2.510-31.000\\
\hline
\label{tab:lik}
\end{tabular}
\end{minipage}
\end{table*}

\begin{table*}
 \centering
 \begin{minipage}{120mm}
 \caption{Same as Table~\ref{tab:lik} but for an $\Omega_M=0.3$, $\Omega_{\Lambda}=0.7$
 universe. }
 \begin{tabular}{@{}ccccccc@{}}
\hline
\hline
      & Spheres  &            &         & Disks   &            &\\
\hline
      &          &            & 95\% &            &            & 95\%\\
      &          &            &Confidence&        &            &Confidence\\
$z_{\rm l}$ & Mode $R$ & Median $R$ & Limits & Mode $R$ & Median $R$ & Limits\\
   &($h_{72}^{-1}$ kpc)&($h_{72}^{-1}$ kpc)&($h_{72}^{-1}$ kpc)
   &($h_{72}^{-1}$ kpc)&($h_{72}^{-1}$ kpc)&($h_{72}^{-1}$ kpc)\\
\hline
0.7000 & 0.870 &  1.300 &  0.500-5.610 &   1.280 & 1.960 &   0.710-8.760\\
1.0620 & 1.240 &  1.860 &  0.720-8.090 &   1.830 & 2.800 &   1.010-12.470\\
1.1727 & 1.360 &  2.040 &  0.800-9.140 &   2.000 & 3.070 &   1.110-14.000\\
1.1810 & 1.360 &  2.050 &  0.800-9.030 &   2.010 & 3.090 &   1.120-14.090\\
2.9740 & 4.630 &  6.950 &  2.690-29.890 &  6.790 & 10.420 &  3.720-44.670\\
\hline
\label{tab:lik2}
\end{tabular}
\end{minipage}
\end{table*}

The results of the likelihood analysis for different lensing redshifts
are summarized in Figure
\ref{zlik}
and Tables~\ref{tab:lik} and \ref{tab:lik2}. The plots show the
differential likelihood distributions for the two assumed geometries,
${\cal L}(R)$, which peak at the most probable radii for given
$z_{\rm l}$. Also plotted are the cumulative distributions from which median
$R$ values have been calculated together with 95\% confidence
intervals. For spherical absorbers in a non-zero $\Lambda$ universe
the most likely sizes ($2R$) are in the range from 1.74 to 9.26
$h^{-1}_{72}$ kpc. Median sizes are in the range from 2.6 to 13.9
$h^{-1}_{72}$ kpc. For randomly inclined disks the most likely sizes
lie between 2.56 and 13.58 $h^{-1}_{72}$ kpc. Median sizes are between
3.92 and 20.84 $h^{-1}_{72}$ kpc. In a zero $\Lambda$ universe these
values are smaller by a factor of $\sim 1.4$.  

\section{Discussion}\label{sec:disc}
We have used a simple model to select candidate lines for which the
ADR is due to coverage of one LOS. Our model only considered two LOS,
the corresponding images to which, nevertheless, are not equally
bright. As in previous work on the subject, only two simple geometries
have been considered for the absorbing structures. These geometric assumptions are simple not only
in terms of the assumed shapes but also in that these
structures more realistically should have no sharp edge: even in the
simplest scenario the column density has a profile which falls smoothly radially
outwards. Thus the absorbers giving rise to lines x15, x28 and x29 may
conceivably have the same column density profile as the rest but they may be
positioned with respect to a LOS in such a way that they present a
different covering factor. However, incorporation of such more
realistic models is only possible when separate spectra for different
LOS are available (e.g. D'~Odorico et al. 1998, Rauch et al. 2001).

The use of simulated spectra to reproduce an ADR is crucial because
the observed ADR by itself does not provide enough information to
establish the partial coverage effect. For example, group (iv) lines
did not show an ADR when simulated. Also the magnitude of the
adjustment does not necessarily constrain the model sufficiently. As
explained, in theory one would expect the adjustment to agree with
$1-f_R$ calculated from the residual intensities (columns 7 and 10 in
\tr{tab:adj}). In practice, this was not always the case. This was
probably due to blending with nearby lines. Blending affects the
observed residual intensities at the velocity position of the line of
interest, leading to an inaccurate value for $1-f_R$. It may also have
an effect on the magnitude of the adjustment performed, leading to
further discrepancy between the values in columns 7 and 10. For
example, line 3 (adjustment~=~$0.54\pm 11$, $1-f_R=0.211566$) was
simulated with an ADR but lies 5.7~\kmps\ from line 2 ($b=12.8$~\kmps,
$\log N=13.07$) and 13.6~\kmps\ from line 7 ($b=18.6$~\kmps, $\log
N=13.53$). Although when simulated and fitted the line was in
excellent agreement with the $\alpha=1.520$ model, for the observed
and fitted line there was some discrepancy (though not within the
error) between actual adjustment and $1-f_F$, and a more serious
discrepancy between actual adjustment and $1-f_R$.  Of course some
degree of agreement is necessary, at least between $1-f_F$ and actual
adjustment. As explained before, this is how we chose the best value
of $\alpha$ for our simulations.

On the other hand, line 17 does not suffer from as much blending, is
stronger and the discrepancy is smaller.

More insight on the threshold Voigt parameters a line needs to have
for partial coverage to lead to an observable ADR, and on the reality
of the effect can be obtained by inspection of group (i)
lines. Specifically, such lines should be below the estimated
threshold mentioned above (\scr{sec:res}), in that one or more of
the following are true: The lines have low
\logn, large \b, or they lie in a region with low SNR. 
Parameters for all group (i) lines with \logn(\cfour)~$\ge 13.0$ are
shown in \tr{tab:group1}. It can be seen that usually lines are too broad
for their column density. For line 93 the problem may be aggravated by
a relatively low SNR.

\begin{table*} 
              
 \centering
 \begin{minipage}{120mm}
 \caption{Lines with \logn(\cfour)~$\ge 13.0$ 
for which an ADR was neither observed in the fitted spectrum nor
generated in the simulations. Symbols have the same meaning as
in previous tables. The last two columns give the average SNR in the
1548 and 1550~\AA\ regions.}
 \begin{tabular}{@{}ccccccccc@{}}
\hline
index& $z$   &$\sigma_z$&  $b$ &$\sigma_b$&\logn& $\sigma_{{\rm log} N}$ & SNR$_{\rm 1548}$ & SNR$_{\rm 1550}$ \\
\hline
12 & 3.107982 & 0.000034 & 24.8 & 10.6 & 13.137 & 0.146& 95 & 95\\
21 & 3.169724 & 0.000011 & 41.3 & 1.4  & 13.439 & 0.010& 75 & 100\\
24 & 3.171289 & 0.000032 & 44.1 & 4.6  & 13.089 & 0.039& 75 & 100\\
41 & 3.384759 & 0.000041 & 38.0 & 2.7  & 13.018 & 0.038& 85 & 85\\
43 & 3.386163 & 0.000011 & 26.9 & 0.9  & 13.534 & 0.015& 85 & 85\\
47 & 3.378988 & 0.000026 & 26.1 & 1.8  & 13.273 & 0.028& 85 & 85\\
48 & 3.379746 & 0.000029 & 22.9 & 2.3  & 13.251 & 0.055& 85 & 85\\
59 & 3.502039 & 0.000000 & 12.6 & 0.4  & 13.010 & 0.014& 90 & 90\\
67 & 3.558324 & 0.000037 & 16.1 & 1.7  & 13.323 & 0.121& 80 & 80\\
68 & 3.558079 & 0.000006 & 6.5  & 1.1  & 13.107 & 0.170& 80 & 80\\
78 & 3.655437 & 0.000074 & 163.4& 129.4& 13.557 & 0.856& 75 & 75\\
80 & 3.668858 & 0.000016 & 35.6 & 2.0  & 13.283 & 0.034& 75 & 75\\
93 & 3.065258 & 0.000002 & 11.4 & 0.3  & 13.255 & 0.009& 60 & 75\\
97 & 2.973627 & 0.000000 & 34.2 & 2.2  & 13.104 & 0.034& 85 & 85\\
98 & 2.974296 & 0.000070 & 39.4 & 4.1  & 13.135 & 0.080& 85 & 85\\
\hline
\label{tab:group1}
\end{tabular}
\end{minipage}
\end{table*}
 
Given the small size of the sample, these results cannot be considered
representative of intervening \cfour\ absorbers in general. Rather
they should be considered as an illustration of what can be learned
from a compilation of similar studies of gravitationally lensed
QSOs. From Equations~\ref{phis} to~\ref{lik} it follows that these
results depend on LOS separations for the sample lines and on lensing
redshifts chosen. However, taken at face value they strongly suggest
that {\it \cfour\ absorption arises in structures with sizes on the
order of kiloparsecs.}  It can be seen from Tables~\ref{tab:lik}
and~\ref{tab:lik2} that, for the three $z_{\rm l}$ values 1.062, 1.1727
and 1.181, estimated sizes are of the order of $\sim 2$
kpc. On the other hand, Tables~\ref{tab:sep} and~\ref{tab:sep2} suggest
that the structures responsible for intervening \cfour\ absorption
cannot be smaller than $\sim 0.2 h^{-1}_{72}$ kpc for
$\Omega_M=1$, $\Omega_{\Lambda}=0$ ($\sim 0.3 h^{-1}_{72}$ kpc for
$\Omega_M=0.3$, $\Omega_{\Lambda}=0.7$). 

Up to this point we have implicitly assumed that the estimated
\lq sizes\rq\ are those of coherent, discrete entities. In particular, the
maximum likelihood method is based on an assumption about the
geometric shape of such objects. However, whether one has separate
spectra for different LOS or not, the same observations can
also be due to an ensemble of correlated objects of smaller
characteristic size. In this case total coverage provides an estimate
for a coherence length of clustered absorbers. Alternatively, there
might be a lack of discrete objects at any scale, with partial
coverage being the effect of variations in an inhomogeneous
density field permeating the IGM.

If total coverage is due to correlated discrete absorbers, then in
practice we may be \lq measuring\rq\ transverse sizes of clumpy systems. A
way to investigate this issue is by observing systems which show
several lines above the ADR detectability threshold and close in
velocity space. If these proved to be {\it p}-type lines, we would
probably be dealing with a clumpy system covering one LOS. We claim
that \fr{fig:f3.10} argues against such a picture for the \zap3.10
system since line 7 ({\it t}) lies between lines 3 and 4 ({\it p}).

There are two main caveats concerning our results. First, the sample
of lines is small and this is expected to affect the estimates
themselves because it is not just the fraction of hits and misses that
is important (Fang et al. 1996). The most probable size calculated
reflects the LOS separation over a small redshift range.  It is
unclear whether the result would remain similar if a broader redshift
range were covered. If such a sample were available, it would be
possible to carry out size estimates in different redshift bins so as
to look for any evolutionary trends in the estimated absorber
sizes. Evolution of the UV background field would be expected to have
an effect on absorber sizes. If these are predominantly photoionized,
we might expect size estimates for \cfour\ absorbers to become smaller
at low \z. Note however that apparent redshift evolution may be an
artefact due to LOS separation (Fang et al. 1996; Dinshaw et al. 1998) and,
therefore, difficult to confirm.

Second, the redshift of the lens is unknown; multiple lensing would
introduce further complications. Thus these results only give an order
of magnitude estimate of \cfour\ absorber size.

It is worth noting that these results agree in order of magnitude, at
least for the lower $z_{\rm l}$ values, with recent results from work
on a larger sample with separate spectra for different LOS (Rauch et
al. 2001).  These authors found that there is very little difference
between column densities in separate spectra for LOS separations below
about $300 \ h_{50}^{-1}$~pc. On the other hand, column density
differences reach 50\%\ just below LOS separations of $\sim 1 \
h_{50}^{-1}$~kpc. From energy arguments these authors conclude that
\cfour\ clouds are the end products of ancient galactic outflows.
There is also independent evidence for such a scenario.  Theuns et
al. (2002) showed that hydrodynamical simulations with feedback
produce \cfour\ absorption lines with properties in reasonable
agreement with observations.  They used the same Voigt profile fits to
\cfour\ lines which we have used in the present paper, together with
fits to one more QSO to demonstrate this result.  It is thus possible
that the structures examined here represent clouds whose baryonic
content has been transported, or at least metal-enriched, to large
distances (up to $\sim 1$~Mpc) from the central regions of their
parent galaxies.

Our results also agree with the results of Petitjean et al. (2000) for
\mgtwo\ systems. These authors used equivalent width models in which
one LOS was saturated and one optically thin to estimate that
\logn(\mgtwo) for $z < 1.7$ absorbers changes by at least an order of
magnitude over less than $\sim 1 \ h_{75}^{-1}$~kpc. However, there is
no indication that the two populations may belong to the same class,
as there are no low ionization lines associated to any of our \cfour\
lines which show an ADR. Although there are some low ionization lines
associated to some of the other \cfour\ lines in the full sample,
these do not show multiplets, so that it is not possible to look for
evidence of partial coverage.  There is also no overlap in redshift
between the two populations.

On the other hand, if our somewhat higher estimates for $z_{\rm
l}=2.974$ are closer to reality, then these would support the picture
according to which intervening \cfour\ absorbers arise in
PGCs (Haehnelt et al. 1996, Rauch et al. 1997).

Note that in total there are 8 (10) components which are either {\it p} 
or {\it t}, depending on whether we count lines 25 and 26.  4 out
of these are {\it p} which is a fraction of 0.5 (0.4). Of course, this
is a single case and its statistical significance cannot be estimated.
However, even if, in a worst case scenario, we trust only one
(observed) line, this result says that the probability of partial
coverage for lines above the detectability limit is not
negligible. This in turn suggests that for gravitationally lensed QSOs 
partial coverage by intervening structures is not rare and observing
it is mainly an issue of detectability.

\section{Conclusions}\label{sec:conc}
The spectra used in this work are the result of
careful wavelength and flux calibration, and order merging.
The observed ADRs are thus real, and we have shown that
in many cases they can be simulated by means of a simple physical model.
From LOS separations at different redshifts we obtained
lower limit estimates of $\sim 300 \ h_{72}^{-1}$~pc for \cfour\ absorbers.
From a statistical approach we obtained most probable sizes of a few kiloparsecs, 
but, because the sample is small, this estimate is rather uncertain.

We have briefly discussed the possible nature of the absorbers based
on our size estimates. Clearly, more work needs to be done before we
can decide among different possibilities. In this respect it would be
interesting to carry out a cross-correlation study between Lyman break
galaxies and our \cfour\ absorbers similar to the work by Adelberger
et al. (2002) who found that the two classes may be related.

Ideally, one would like to have separate spectra for different
LOS. However, given that lensed QSOs may have small image separations,
it is difficult to find suitable candidates. The technique used here
compensates for this problem as it allows probing of closer LOS. All
that is required is the existence of two images with a known angular
separation as well as an estimate for the redshift of the lens. As in
the case studied here, the separation may be obtained in the course of
different observations with different instruments (e.g. space based)
or in different wavebands (e.g. radio) where the necessary resolution
can be attained.  Obtaining a reliable value for $z_l$ is more
difficult as it depends on a combination of modelling and detection of
the lensing mass, but if there are several candidates, several models
can be run.

In principle the method could be extended to more than two
images. However, with several images of different brightnesses the
uncertainties in the zero level adjustment parameter would probably
allow different configurations. It may be possible to add a
probability argument for these as well, but adding this degree of
complexity for what would be a small sample is not likely to yield
worthwhile results.

Finally, it is worth noting that ADRs can be useful in a different way
as well.  In cases of spectra of QSOs which are not known to be
lensed, the need for same, or similar, ADR corrections for several
absorbers would be a strong indication that a QSO is lensed. The
method illustrated in this paper would then have to be modified to
include the angular separation as an unknown as well, with more
speculative results.

\section*{Acknowledgments}
The observations were made at the W. M. Keck Observatory,
which is operated as a scientific partnership between the California
Institute of Technology and the University of California; it was made
possible by the generous support of the W. M. Keck Foundation.

\label{lastpage}

\end{document}